\documentclass[conference]{IEEEtran}
\IEEEoverridecommandlockouts

\scrollmode
\usepackage{amsmath}
\usepackage{amsfonts}
\usepackage{amssymb}
\usepackage{latexsym}
\usepackage{stmaryrd}
\usepackage{array}
\usepackage{exscale}
\newcommand{\nc}{\newcommand}

\newcommand{\ve}{\varepsilon}
\newcommand{\vp}{\varphi}

\newcommand{\sse}{\subseteq}

\newcommand{\spe}{\supseteq}
\newcommand{\fa}{\forall}
\newcommand{\ex}{\exists}
\newcommand{\mr}{\mathrm}
\newcommand{\mc}{\mathcal}
\newcommand{\mf}{\mathfrak}

\newcommand{\DMO}{\DeclareMathOperator}
\newcommand{\DST}{\displaystyle}

\newcommand{\RR}{\mathbb{R}}

\mathchardef\breakingcomma\mathcode`\,
{\catcode`,=\active
  \gdef,{\breakingcomma\discretionary{}{}{}}
}

\usepackage{listings}
\newcommand{\inl}[1]{\lstinline$#1$}
\nc{\simlvi}[1]{\!\sim_{#1}}
\nc{\apprel}[3]{{#1}(#2)_{(#3)}} 
\nc{\cmpli}[1]{\complement^1_{#1}} 
\nc{\cmplzi}[1]{\complement^0_{#1}} 
\nc{\cmplzoi}[1]{\complement^*_{#1}} 

\nc{\zf}{\mr{ZF}}
\nc{\zfmf}{\zf^0} 
\nc{\zfc}{\mr{ZFC}}
\nc{\zfcmf}{\zfc^0} 
\nc{\bst}{\mr{BST}} 
\providecommand{\norm}[1]{\lVert #1 \rVert} 
\makeatletter
\DeclareRobustCommand{\genericinterval}[2]{%
  \@ifstar{\genericinterval@star{#1}{#2}}{\genericinterval@nostar{#1}{#2}}}
\newcommand{\genericinterval@star}[4]{\mathopen{}\mathclose{\left#1#3,#4\right#2}}
\newcommand{\genericinterval@nostar}[4]{\mathopen{#1}#3,#4\mathclose{#2}}

\makeatother
\nc{\untit}[2]{{#1}^{#2 \downarrow}} 
\nc{\obit}[2]{{#1}^{#2 \uparrow}} 
\nc{\inzEKi}[1]{\mc{I}^{\mr{V}}_{#1}}

\nc{\inzKEi}[1]{\mc{I}^{\mr{E}}_{#1}}

\nc{\adjEi}[1]{\mc{A}^{\mr{V}}_{#1}}

\nc{\BD}[1]{{#1}\text{-}\mr{BD}}

\nc{\konv}[2]{{#1}[{#2}]} 
\nc{\actpres}[1]{\Phi_{#1}} 
\nc{\Prim}{\mc{PR}} 

\nc{\sselr}{\sse^{\mapsto}}
\nc{\sserl}{\sse^{\mapsfrom}}
\nc{\spelr}{\spe^{\mapsto}}
\nc{\sperl}{\spe^{\mapsfrom}}
\nc{\ball}[1]{\mr{B}^{#1}} 
\nc{\oball}[1]{\breve{\mr{B}}^{#1}} 
\nc{\pball}[1]{\dot{\mr{B}}^{#1}} 
\nc{\prr}[1]{\dot{\RR}^{#1}} 
\nc{\sph}[1]{\mr{S}^{#1}} 
\nc{\ssim}[1]{s\sigma_{#1}} 
\nc{\koerper}[1]{\norm{#1}}
\nc{\Ccovdim}{\mc{CD}}
\nc{\Cinddim}{\mc{SID}}

\nc{\CInddim}{\mc{LID}}

\DeclareMathOperator{\diffop}{D} 
\DeclareMathOperator*{\diffoplimit}{D} 
\nc{\diffopc}[1]{\sideset{_{#1}}{}\diffoplimit} 
\nc{\diffopp}[1]{\diffop_{#1}} 
\nc{\diffopcp}[2]{\sideset{_{#2}}{_{#1}}\diffoplimit} 
\nc{\meanH}[2]{\mf{M}_{#1,#2}} 
\nc{\emean}[2]{\mf{M}_{\exp_{#1},#2}} 
\DeclareMathOperator{\mor}{Mor}
\DeclareMathOperator{\Hom}{Hom} 
\nc{\autoerw}[1]{\mr{Aut}^{#1}} 
\nc{\komma}[2]{(#1 \downarrow #2)} 
\nc{\Kmat}{\mf{MAT}} 
\nc{\Khmat}{\mf{HMAT}} 
\nc{\homfun}[1]{\mor_{#1}(-_1,-_2)} 
\nc{\homfunae}[1]{\mor_{#1}(-_1)} 
\nc{\homfunbe}[1]{\mor_{#1}(-_2)} 
\nc{\homfunxy}[3]{\mor_{#1}(#2(-_1), #3(-_2))}
\nc{\homfunx}[2]{\mor_{#1}(#2(-_1), -_2)}
\nc{\homfuny}[2]{\mor_{#1}(-_1, #2(-_2))}
\nc{\homfuna}[2]{\mor_{#1}(#2, -)} 
\nc{\homfunb}[2]{\mor_{#1}(-, #2)} 
\nc{\hhomfuna}[2]{\Hom_{#1}(#2, -)} 
\nc{\hhomfunb}[2]{\Hom_{#1}(-, #2)} 

\newcommand{\Cls}{\mc{CLS}}

\newcommand{\Musat}{\mc{M\hspace{0.8pt}U}} 
\newcommand{\Musati}[1]{\Musat_{\!#1}} 
\newcommand{\Smusat}{\mc{S}\Musat} 
\newcommand{\Smusati}[1]{\Smusat_{\!#1}}

\nc{\Clsoo}{\Cls^{1,1}} 

\DeclareMathOperator{\var}{var}

\DMO{\dos}{ds} 
\DMO{\mdos}{mds} 
\newcommand{\Clash}{\mc{HIT}} 

\newcommand{\Uclash}{\mc{U}\Clash} 
\newcommand{\Uclashi}[1]{\Uclash_{\!\!#1}}
\DMO{\premr}{ax} 
\DMO{\concr}{C} 
\DMO{\allcr}{cl} 

\DMO{\thardness}{thd} 
\DMO{\phardness}{phd} 
\DMO{\whardness}{awid} 
\DMO{\dep}{dep} 
\DMO{\hts}{hs} 
\DMO{\semspace}{css} 
\DMO{\resspace}{crs} 
\DMO{\treespace}{cts} 
\nc{\bth}[1]{\langle{#1}\rangle} 
\DMO{\rsub}{r_S} 
\DMO{\rk}{r} 
\DMO{\ro}{\rk_1} 
\DMO{\rki}{\rk_{\infty}} 
\DMO{\rpl}{r^{pl}} 
\DMO{\ropl}{\rk_1^{pl}} 
\nc{\rslur}{\xrightarrow{\text{SLUR}}} 
\nc{\rslurs}{\rslur_{\!*}} 
\DMO{\slur}{slur} 
\nc{\Slur}{\mc{SLUR}} 
\nc{\rkslur}[1]{\xrightarrow{\text{SLUR}_{#1}}} 
\nc{\rkslurs}[1]{\rkslur{#1}_{\!*}} 
\nc{\Altsluri}[1]{\Slur(#1)}
\nc{\Altslurstari}[1]{\Slur\text{\textasteriskcentered}(#1)}
\nc{\Canoni}[1]{\mr{CANON}(#1)}
\nc{\rkslurstar}[1]{\xrightarrow{\text{SLUR\textasteriskcentered}#1}} 
\nc{\rkslursstar}[1]{\rkslurstar{#1}_{\!*}} 
\DMO{\slurstar}{\slur\!\text{\textasteriskcentered}}
\nc{\Urefc}{\mc{UC}}
\nc{\Propc}{\mc{PC}}
\nc{\Wrefc}{\mc{WC}} 
\DeclareMathOperator{\vdeg}{vd} 
\DeclareMathOperator{\minvdeg}{\mu\!\vdeg} 
\DMO{\varmvd}{\var_{\minvdeg}} 
\DMO{\nfc}{fc} 
\DMO{\maxnfc}{\nu\!\nfc} 
\nc{\Dt}[1]{\mc{F}_{#1}} 

\nc{\svbf}{\mc{VB}} 
\nc{\svbfs}{\mc{VB}^*} 
\DMO{\potp}{pp} 
\DMO{\potprec}{NM} 
\DMO{\minnonmer}{VDM} 
\DMO{\minnonmerh}{VDH} 
\DMO{\maxsmar}{FCM} 
\DMO{\maxsmarh}{FCH} 
\DMO{\varsing}{\var_s} 
\DMO{\varosing}{\var_{1s}} 
\DMO{\varnosing}{\var_{\neg1s}} 
\DMO{\nsv}{\mathit{n}_s} 
\DMO{\nosv}{\mathit{n}_{1s}}
\DMO{\nnosv}{\mathit{n}_{\neg1s}}
\nc{\Musatns}{\Musat'} 
\nc{\Musatnsi}[1]{\Musati{#1}'}
\nc{\Smusatns}{\Smusat'} 
\nc{\Smusatnsi}[1]{\Smusati{#1}'}
\nc{\Uclashns}{\Uclash'} 
\nc{\Uclashnsi}[1]{\Uclashi{#1}'}
\nc{\tsdp}{\xrightarrow{\text{sDP}}}
\nc{\tsdps}{\tsdp_{\!*}}
\nc{\tosdp}{\xrightarrow{\text{1sDP}}}
\nc{\tosdps}{\tosdp_{\!*}}
\DMO{\sdp}{sDP} 
\DMO{\osdp}{sDP_1} 
\nc{\cflmusat}{\mc{CF}\Musat} 
\nc{\cflmusati}[1]{\mc{CF}\Musati{#1}}
\nc{\cflimusat}{\mc{CFI}\Musat} 
\DMO{\sNF}{sNF} 
\DMO{\eqp}{eqp} 
\DMO{\sgp}{sp} 
\DMO{\singind}{si} 
\DMO{\osingind}{si_1} 
\DMO{\shyp}{svh} 
\DMO{\sdph}{ssh} 
\DMO{\msdph}{mss} 
\DMO{\osdph}{ssh_1} 
\DMO{\mosdph}{mss_1} 
\DMO{\mps}{mps} 
\DMO{\purec}{puc} 
\DMO{\doping}{D}
\nc{\glue}[4]{\operatorname{glue}((#1,#2), (#3,#4))} 
\nc{\gluea}[3]{#1 \mathbin{\boxplus}_{#3} #2} 
\DMO{\saturate}{S}
\DMO{\marginalise}{M}
\DMO{\frl}{fl} 
\nc{\Con}{\mr{Con}}
\nc{\Log}{\mr{Log}}
\nc{\Lin}{\mr{Lin}}
\nc{\Pol}{\mr{Pol}}
\nc{\ExL}{\mr{ExL}}
\nc{\ExP}{\mr{ExP}}
\nc{\CTime}{\mr{CTime}}
\nc{\CSpace}{\mr{CSpace}}
\nc{\LTime}{\mr{LTime}}
\nc{\LSpace}{\mr{L}}
\nc{\NLSpace}{\mr{NL}}
\nc{\LinTime}{\mr{LinTime}}
\nc{\LinSpace}{\mr{LinSpace}}
\nc{\PTime}{\mr{P}}
\nc{\PSpace}{\mr{PSpace}}
\nc{\Np}{\mr{NP}}
\nc{\Conp}{\text{coNP}}
\nc{\NPSpace}{\mr{NPSpace}}
\nc{\CoNPSpace}{\mr{coNPSpace}}
\nc{\ELTime}{\mr{ELTime}}
\nc{\ELSpace}{\mr{ELSpace}}
\nc{\EPTime}{\mr{EPTime}}
\nc{\EPSpace}{\mr{EPSpace}}
\nc{\NEPTime}{\mr{NEPTime}}
\nc{\polydelta}[1]{\Delta_{#1}^{\mr P}}
\nc{\polypi}[1]{\Pi_{#1}^{\mr P}}
\nc{\polysigma}[1]{\Sigma_{#1}^{\mr P}}
\nc{\Ph}{\mr{PH}}

\nc{\Dp}{D^P}
\nc{\PllC}[2]{{\text{$\mr{PT}$/$\mr{WK}$}(#1, #2)}} 
\nc{\Nc}{\mr{NC}}
\nc{\Nci}[1]{\Nc^{#1}}
\nc{\Ac}{\mr{AC}}
\nc{\Aci}[1]{\Ac^{#1}}
\nc{\pmodpoly}{P / \mathrm{poly}}
\nc{\Wh}[1]{\mr{W}[#1]} 
\nc{\Rl}{\mr{RL}}
\nc{\coRl}{\mr{coRL}}
\nc{\Rp}{\mr{RP}}
\nc{\coRp}{\mr{coRP}}
\nc{\Zpp}{\mr{ZPP}}
\nc{\Bpp}{\mr{BPP}}
\nc{\Pp}{\mr{PP}}
\nc{\Reach}{\mr{STCON}} 
\nc{\Undreach}{\mr{USTCON}} 
\nc{\Pcol}[2]{\mr{COL}(#1,#2)} 
\nc{\Pscol}[2]{\mr{SCOL}(#1,#2)} 
\nc{\Psorcol}[2]{\mr{SORCOL}(#1,#2)} 
\DMO{\slp}{slp}
\nc{\Mss}{\mr{MSS}}
\nc{\Key}{\mr{KEY}}
\nc{\Keyi}[1]{\Key_{\!#1}}
\nc{\Nbmss}{N_{\mr{bm}}} 
\nc{\Nbkey}{N_{\mr{bk}}} 
\nc{\Rnb}{N_{\mr{b}}}
\nc{\Rnk}{N_{\mr{k}}}
\nc{\Rnr}{N_{\mr{r}}}

\nc{\Byte}{\mr{B}[8]}
\nc{\QByte}{\mr{B}[4,8]}
\nc{\KByte}{\mc{B}} 
\nc{\RQByte}{\mc{QB}} 

\nc{\ramz}[3]{\mr{ram}_{#1}^{#2}(#3)} 
\nc{\waez}[2]{\mr{vdw}_{#1}(#2)} 
\nc{\gtz}[2]{\mr{grt}_{#1}(#2)} 
\nc{\pdwaez}[2]{\mr{vdw}_{#1}^{\mr{pd}}(#2)} 
\nc{\absfeh}[1]{\delta_{#1}} 
\nc{\relfeh}[1]{\ve_{#1}} 
\usepackage{theorem} 
\usepackage[driverfallback=hypertex]{hyperref} 
\newtheorem{defi}{Definition}[section]
\newtheorem{lem}[defi]{Lemma}

\theorembodyfont{\rmfamily}

\theorembodyfont{}
\newcounter{dDef} 

\newcounter{dLem} 

\newcounter{dThm} 

\newcounter{dPro} 

\newcounter{Beispielzaehler}

\nc{\bm}{\boldmath}
\nc{\bmm}[1]{\mbox{\bm$\DST #1$}}
\nc{\mi}[1]{\bmm{\mathrm{(#1):}} \quad}
\usepackage{enumerate}
\usepackage[all]{xy}
\usepackage[active]{srcltx}

\DeclareMathOperator{\Aaut}{A}
\DeclareMathOperator{\Eaut}{E}

\begin{document}

\title{Improving Reasoning on DQBF}

\author{
\IEEEauthorblockN{Ankit Shukla}
\IEEEauthorblockA{
Johannes Kepler University, Austria\\
Email: ankit.shukla@jku.at}
}

\maketitle

\begin{abstract}
The aim of this PhD project is to develop fast and robust reasoning tools for dependency quantified Boolean formulas (DQBF). 
In this paper, we outline two properties, autarkies and symmetries, that potentially can be exploited for pre- and in-processing in the DQBF solving process.
%
DQBF extend quantified Boolean formulas (QBF) with non-linear dependencies between the quantified variables.
Automated testing and debugging techniques are an essential part of the solvers tool-chain.
%
%
For rigorous DQBF solver development, we are working on novel automated testing and debugging techniques as successfully established in SAT and QBF solving. 
%
%
The tool is under development.   
\end{abstract}

\section{Introduction}
\label{sec:Intro}
Dependency Quantified Boolean Formulas (DQBF) are an extension of QBF
which allows the specification of non-linear dependencies between quantified variables. 
For example \par\noindent
\begin{equation}
\begin{split}
&F := \fa x_1,x_2,x_3 \ex y_1(x_1,x_2) \ex y_2(x_2,x_3) \ex y_3(x_1): F_0 \\
&F_0 := (y_1 \vee x_1) \wedge (\overline{y_1} \vee x_2) \wedge (\overline{y_2} \vee \overline{x_2} \vee x_3) \\
&\hspace{2.3cm}\wedge (y_3 \vee \overline{x_1} \vee x_2) \wedge (\overline{y_3} \vee x_1)
\end{split}
\label{eqn:ex}
\end{equation}

is a DQBF formula where explicit dependencies ($(x_1,x_2), (x_2,x_3), (x_1)$) of quantified existential variables ($y_1, y_2, y_3$ respectively) are specified. We call the explicit dependencies the \textit{dependency set} of the corresponding variable. 
The rest of the formula $F_0$ is the propositional formula in the conjunctive normal form (CNF).
DQBF (in general) can offer more succinct descriptions than
QBF.
Deciding DQBF is NEXPTIME-complete, compared to the PSPACE completeness of QBF.
Making it suitable to model problems known to be NEXPTIME-complete, for e.g. partial information non-cooperative games, program and circuit synthesis, probabilistic planning of finite length, and certain bit-vector logics.

\section{preprocessing}
\label{sec:pre}
Preprocessing techniques~\cite{een2005effective} tries to reduce the input formula by simplification procedures before the formula is passed to the actual solving algorithm.
It is well acknowledged by the SAT and QBF community that these preprocessing techniques often reduce the computation time of the solver by orders of magnitude.
Inprocessing, on the other hand, uses the formula simplification procedures during the search process of the solver. 
For DQBF in practice, it might pay off to spend
more time on pre/inprocessing due to the hardness of the problem.

We propose to use autarkies and symmetries for pre- and inprocessing as discussed in the following. 
%
%

%
%

\subsection{Autarkies}
\label{sec:aut}
An autarky for a CNF $F$ is a partial assignment (mapping variables of the formula to \textit{true} or \textit{false}) which either do not ``touch" a clause (no variable of that clause is assigned) or satisfies it.
Clauses satisfied by some autarky can be removed satisfiability-equivalently. 
We generalize the concept for DQBF by considering partial assignments to existential variables and allowing boolean functions of universal variables as values (fulfilling the dependencies) substituted for them.
Now the clauses with assigned variables need to become tautologies.
Note that an empty partial assignment is an autarky for every $F$ i.e. never touching any clause (\textit{trivial autarky}) and a satisfying assignment for $F$ is also an autarky for $F$, touching every clause and satisfying it.

A DQBF is called \textbf{lean} if it has no non-trivial autarkies.
The union of two lean DQCNF with compatible variables and dependencies is again lean, and thus every DQBF has a largest lean sub-DQBF, the \textbf{lean kernel}.
Alternatively one can arrive at the lean kernel via \textbf{autarky reduction}.
We denote by $F[\phi]$ the DQBF with the clauses removed which are satisfied by an autarky $\phi$. \vspace{-1em}
\begin{lem}\label{lem:autsateq}
	$F[\phi]$ is satisfiability-equivalent to $F$ for an autarky $\vp$ of $F$.
\end{lem}
\begin{lem}\label{lem:compaut}
Autarky reduction is confluent.
\end{lem}

For two autarkies $\vp, \psi$ of $F$ one can consider the composition, 
 which on the variables of $\psi$ acts like $\psi$, and otherwise like $\vp$: \vspace{-1em}
\begin{lem}\label{lem:compaut}
	The composition of two autarkies is again an autarky.
\end{lem}
Now the lean kernel is obtained by repeatedly applying autarky-reduction on $F$ as long as possible:
\vspace{-1em}
\begin{lem}\label{lem:decomp}
	Consider a DQBF $F$. The largest lean sub-DQBF is also obtained by applying autarky-reduction to $F$ as long as possible (in any order).
\end{lem}

We present two types of autarky systems for DQBF, namely, $A$ and $E$ (using ``A" to denote universal variables, and ``E" for existential variables).
Consider a DQCNF $F$ and $k \ge 0$:
\begin{itemize}
	\item An \textbf{$\Aaut_k$-autarky} for $F$ is an autarky such that all boolean functions assigned depend essentially on at most $k$ variables.
	\item An \textbf{$\Eaut_k$-autarky} is an autarky assigns at most $k$ (existential) variables.
\end{itemize}

$A_0, A_1$ allow the boolean functions to essentially depend on 0 resp. 1 universal variable, while $E_1$ only uses one existential variable (for a single autarky).
Deciding the existence of $E_1$-autarky can be done in polynomial time.
whereas for $A_1$-autarky it is NP-complete and require a SAT solver.

Consider the Example~\ref{eqn:ex}.
Since $\overline{y_2}$ is pure, we have the $\Aaut_0$-autarky $y_2 \rightarrow 0$ (removing the third clause).
Furthermore we have the $\Aaut_1$-autarky $y_3 \rightarrow x_1$, removing the fourth and fifth clauses.
Both these autarkies are also $\Eaut_1$-autarkies.
We obtain the reduction result $\fa x_1,x_2,x_3 \ex y_1(x_1,x_2) : (y_1 \vee x_1) \wedge (\overline{y_1} \vee x_2)$, 
which doesn't allow any further autarky. 
The remaining formula is the lean kernel of the original formula.
We can now try to solve this reduced formula using a DQBF solver.

As demonstrated in the above example we can use autarky procedure to simplify the input formula, but in practice, the autarky system $A_0, A_1$ and $E_1$ are too weak for the general DQBF solving. Out of total 334 instances of DQBF track of QBFEVAL'18 we only found 4 instances with non-trivial $A_1$+$E_1$ autakies. We are analyzing stronger autarky systems, $A_2$ (function is dependent on 2 universal variables), $E_2$ (only uses two existential variables) and their combination $E_2 + A_2$. 
Another exciting direction is to use autarky procedure during search (inprocessing) of a solver. 

\subsection{Symmetry}
\label{sec:sym}

%
%
In SAT, symmetries in the formula can often be exploited to speed up  the solving process, because it can prevent a solver from needlessly exploring equivalent parts of a search space.
One way to eliminate symmetries is to add symmetry breaking formulas $\Phi$ (symmetry breaker) to the input formula $F$ known as static symmetry breaking.
%
%
%
%
%

In~\cite{kauers2018symmetries} a general framework for symmetry breaking for QBF was presented. To prove the soundness of the symmetry breaker syntactic restrictions of using variables from the same quantification blocks were enforced.
Similarly, for DQBF we use the following restriction:
\begin{itemize}
	\item The propositional structure is preserved.
	\item The dependency of the variable is the same. 
\end{itemize}

We are currently analyzing less restricted cases where we relax the requirement of the variables to have the same dependency. 

%


\subsection{Synergy of autarkies and symmetry}

The process of finding $A_{1}$-autarkies requires compiling the minimum possibility for each clause to make it a tautology.
For example, to make the first clause in Example~\ref{eqn:ex} $(y_1 \lor x_1)$ a tautology, the minimum possibility is \{$y_1 \to 1$, $y_1 \to \neg x_1$\} (both of these assignments makes the clause a tautology).
Currently, the compilation is clausewise and fed directly to a SAT solver.
We can use the concept of symmetry to reduce the work by a SAT solver and (most likely) improve the efficiency.
The idea of \textbf{symmetry aware compilation} is to make the process of solving the problem with communication across the clauses (at multi-clause level rather than the current state of doing it clausewise).
%
%
Note that the use of symmetry here is similar to finding symmetry during search (dynamic symmetry breaking), unlike the use of the symmetry breaker (static symmetry breaking). 
 

\section{DQBF solver development}
\label{sec:dev}

Robustness and correctness are admissible properties of SAT and
QBF solver implementations.
Tool support for the testing and delta debugging~\cite{brummayer2010automated} are an essential part of the solver ecosystem.
Similar toolsets are necessary for DQBF to make the process of DQBF solving more reliable.  
We identify three critical areas where more development (theory or tooling) is needed.

Fuzzing is an automated testing technique that treats software as a black-box and repeatedly “attacks” it with random inputs to find critical defects.
For this purpose, we establish a new random model for DQBF with interesting theoretical properties like phase transition.
Our main objective is to obtain high code coverage.
We are in the final stages of the development of the tool. 

Delta Debugging is a methodology to automate the debugging of programs to find failure-inducing circumstances. 
If an input caused the solver to crash or returns an incorrect result a delta debugger tries to systematically narrow down the particular input until a given time limit is reached or no further narrowing is possible.
The reduced input can then be used for effective manual debugging.
We are in the early stages of the design for a delta debugger for DQBF.

Testing, however, can never guarantee the correctness of a solver.
Therefore we suggest certifying the solving results by producing proofs that are easy to check as it is standard in SAT + QBF solving in terms of QRAT proofs.
It is also expected that such proof can serve a basis for the extraction of a solution. 
For example, applications of DQBF like program/hardware synthesis, game scenarios, and planning require the extraction of winning strategies or a plan in addition to whether there exist or not answer.
%
%


\section{Conclusion}
\label{sec:conc}

We consider auatarkies and symmetry for pre- and inprocessing of DQBF solving.
We outline the search-based implementation of symmetry during autarky search.
As part of our current and future work for the correct and rigorous development of DQBF solving, we sketch an outline for fuzzing, delta debugging and certification for DQBF.

\bibliographystyle{plainurl}
\bibliography{fmcad}


\end{document}